\documentclass[10pt,twocolumn,letterpaper]{article}
\usepackage[margin=1in]{geometry}
\usepackage{mathptmx}
\usepackage{amsmath,amssymb,booktabs,graphicx,mathtools}
\graphicspath{{figs/}{figures/}{../figures/}}
\usepackage{tikz}
\usetikzlibrary{arrows.meta,patterns}
\usepackage[labelfont=bf,font=small]{caption}
\usepackage{fancyhdr}
\usepackage{url}
\usepackage{enumitem}
\setlist{nosep,topsep=2pt,partopsep=0pt,parsep=0pt,itemsep=1pt,
         leftmargin=1.2em}
\usepackage{titlesec}
\titlespacing*{\section}{0pt}{5pt plus 1pt minus 1pt}{2pt}
\titlespacing*{\subsection}{0pt}{4pt plus 1pt minus 0.5pt}{1.5pt}
\titlespacing*{\subsubsection}{0pt}{3pt plus 0.5pt minus 0.5pt}{1pt}
\renewcommand{\arraystretch}{0.95}
\makeatletter
\renewcommand\section{\@startsection{section}{1}{\z@}
  {-2.4ex \@plus -0.3ex \@minus -.2ex}{1.1ex \@plus .1ex}
  {\normalfont\large\bfseries}}
\renewcommand\subsection{\@startsection{subsection}{2}{\z@}
  {-2.0ex \@plus -0.2ex \@minus -.2ex}{0.8ex \@plus .1ex}
  {\normalfont\normalsize\bfseries}}
\makeatother
\begin{document}
\twocolumn[
  \begin{@twocolumnfalse}
  \begin{center}
  {\large\bfseries IAC-26-C1.3.2\par\vspace{0.4em}
   Decentralized Formation Control with Differential Drag
   for Close-Proximity Satellite Clusters\par}
  \vspace{0.8em}
  {\normalsize P.~McKeen$^{a\ast}$\quad Z.~Manchester$^{a}$\par}
  \vspace{0.3em}
  {\footnotesize\itshape $^{a}$Robotic Exploration Laboratory
   (RExLab), Massachusetts Institute of Technology,
   Cambridge, MA 02139, USA\par
   $^{\ast}$Corresponding author: pmckeen@mit.edu\par}
  \end{center}
  \vspace{-0.6em}
  \begin{center}Abstract\end{center}
  \vspace{-0.5em}
  {\noindent\small Recent proposals for space-based AI infrastructure
envision close-proximity satellite clusters linked by optical
communications for distributed machine learning. Building on
the $81$-satellite concept of Ag\"uera y Arcas et al.\
(2025), at $650$~km in sun-synchronous orbit with
$100$--$200$~m separations, we develop a stationkeeping
budget, passive-safety characterization, and decentralized
control architecture. Each satellite regulates six relative orbital elements
about a fleet-centroid reference using attitude-modulated
forces and scheduled normal thrust. Neighbor-to-neighbor
centroid estimation is assessed separately from control.
 
\noindent Nonlinear simulations with $J_2$, atmospheric drag, and
solar radiation pressure show that attitude modulation
provides in-plane formation-keeping without thrust while
remaining near Sun-pointing. Orbit-normal electric propulsion
maintains the inclination pattern. Across the sampled Sun-angle and density envelope, the controller
holds the $81$-satellite formation at $100$~m spacing to approximately
$3$~m RMS slot error; a $600$-orbit case shows no secular growth. Small slot errors can
conceal deteriorating separation as the eccentricity and
inclination components of the lattice rotate apart. Rotating them together preserves the geometry but roughly
doubles inclination propellant. Including reboost and the
adopted approximations for other uses of thrust, the baseline grid's three-year propellant
budget for a rim (fleet-worst) satellite is $2.3$--$3.8$~kg (96--162 m/s) across fixed
solar-minimum to solar-maximum scenarios. Estimated fleet-mean budgets across all conditions and geometries range from 9.1–170 m/s, or 0.21–4.0 kg.

\noindent Monte Carlo analysis compares formation geometries under
single-vehicle and fleet-wide control loss. At a $35$~m
keep-out, $99\%$ of baseline-grid trials remain clear for
$1.6$~days after one vehicle loses formation control while
retaining its attitude, versus $6.5$~days after fleet-wide
loss. Geometry changes both this response window and its
propellant cost: anisotropic grids retain comparable survival
under no-impulse failures with about $13\%$ less inclination
propellant, while a linear node ladder offers longer survival
at higher cost and lower capacity.

\noindent
Formation geometry sets how much time a cluster has to respond to failure and what that protection costs. Together these results give the concept a controller, a response-time requirement, and a propellant budget, and identify the geometric decisions that trade between them.
\par}
  \vspace{1em}
  \end{@twocolumnfalse}
]

\section{Introduction}

Recent proposals for space-based computing distribute
machine-learning workloads across satellite clusters linked
by free-space optical communications. This trades the
difficulty of building one large orbital server for that of
flying many smaller ones close together. One such
concept~\cite{concept} places $81$ satellites at $650$~km in
a sun-synchronous orbit (SSO), with inter-satellite separations of
$100$--$200$~m, and characterizes their free-fall dynamics.
This paper develops the guidance, navigation, and control
analysis needed to assess formation-keeping feasibility.

Dedicated optical links require assigned neighbors
and slots, and the satellites must be controlled to keep those places while avoiding collision. These satellites have comparatively large area ($105$~m$^2$ broadside area at
$575$~kg), allowing much of this control to be done with environmental forces. Differential drag
and solar radiation pressure provide useful control
authority, reducing required fuel and lowering the threat of a satellite's thrust plumes impinging on its nearby neighbors.

These forces provide in-plane control through
attitude modulation, but maintaining the inclination pattern
against differential $J_2$ precession requires orbit-normal
thrust in most cases studied. Inclination spread also separates the formation
from a failed vehicle, so the same geometric parameter sets
both the passive safety margin and the propellant cost. We examine control, failure response, and geometry
together. The tracked reference is part of this choice:
keeping the $\mathbf{\delta e}$ and $\mathbf{\delta i}$ patterns aligned maintains separation at
about twice the inclination propellant of a fixed-inclination-pattern
reference (Sec.~\ref{sec:refsafety}).

The analysis develops four results:
\begin{itemize}
\item A decentralized control law (Sec.~\ref{sec:control})
  regulates six relative orbital elements about a
  fleet-centroid reference, using attitude-modulated drag
  and solar radiation pressure with thrust confined to the
  orbit normal. It holds an $81$-vehicle lattice at $100$~m
  spacing to a few meters across the operating envelope.
\item A passive-safety characterization (Sec.~\ref{sec:safety})
  uses survival curves against detection latency to understand the collision risks in these formations. We find that the response window at 99\% survival is highly dependent on fault mode and formation geometry (Fig.~\ref{fig:faultpanel}): between 2 hours and 5 days for a single satellite failure (consistently $\approx20$ min with a 5 cm/s velocity error), but between 2 hours and 20 days for a fleet-wide loss.
\item A propellant budget (Sec.~\ref{sec:dv}) separates the
  reboost and inclination maintenance costs, with the geometry-dependent
  inclination cost derived analytically and checked against
  nonlinear simulation. For the isotropic grid, the three-year mean-solar budget is 78 m/s per satellite at the fleet mean and 108 m/s at the rim.
\item A design frontier (Sec.~\ref{sec:trade}) trades passive
  safety against propellant between the $c_x=0$ and isotropic
  endpoints, identifying an anisotropic candidate band. The tested planar configuration has a very low propellant usage but a high risk of collision, while a 1D line in the nodal coordinate offers longer survival at higher propellant cost and with different capacity and network topology.
\end{itemize}
\noindent Together these take the concept from a characterized
free-fall formation to a controlled one with a stated safety
margin and a stated propellant cost, and identify the
geometric decisions as a design trade.

\section{Background and Problem Setup}
\label{sec:background}

\subsection{Prior work}

Differential-drag formation control dates to Leonard's work
in the 1980s~\cite{leonard}. It has flown
operationally~\cite{planet}, generally as an along-track
actuator. Later extensions cover three-dimensional lift and drag under $J_2$~\cite{traub}.

The cited cases are at separations of
kilometers and above, and/or at vehicle counts in the ones and
tens. The present work addresses about $100$ vehicles with roughly $100$~m spacing,
with two actuation mechanisms whose relative contribution
depends on the Sun angle in SSO.

Arrangements that are passively safe are of particular
interest in formation flying, a strong example being eccentricity/inclination
vector separation~\cite{damico}, which the present work
takes as its starting point for the geometry. Koenig and
D'Amico~\cite{koenig18} extend that construction to
$N$-spacecraft swarms in relative orbital elements, with
linear passive-safety constraints and along-track drag
control against differential drag. The present work differs
in separation scale, in the use of attitude-modulated drag
and radiation pressure as the in-plane actuator, and in
treating the inclination spread as a design variable with a
propellant cost.

Cluster keeping by relative-element feedback with
differential drag, including a stability analysis, is
treated by Ben-Yaacov and Gurfil~\cite{benyaacov}, and
multi-spacecraft rendezvous under $J_2$ by Bevilacqua and
Romano~\cite{bevilacqua}. The closest recent result~\cite{bocci} proves asymptotic
stability by Lyapunov methods for a coupled orbit--attitude
drag controller using Hill--Clohessy--Wiltshire dynamics
without $J_2$, constant density, and a prescribed virtual leader. Here, six states are regulated with a two-dimensional saturated command using drag and SRP, and the $J_2$ dynamics determine the lattice trade. These differences prevent direct application of that proof.

\subsection{Constellation baseline}

The baseline is Google's Project Suncatcher concept~\cite{concept}, with capacity for $81$ satellites in a compact, co-orbiting footprint connected by optical links (our ``planar'' configuration). That concept anticipates part of the problem
addressed here. It notes that the cluster's shape is
deformed over the course of an orbit by $J_2$, and proposes
a small retuning of the relative ellipse's axis ratio to
restore periodicity. That correction accounts for the within-orbit deformation. Other geometries explored in this work have secular deformations, discussed in Sec.~\ref{sec:evolution}. Following \cite{concept}, we have chosen Starlink V2 mini as the basis for our physical and geometric model, with assumed parameters given in Table~\ref{tab:vehicle}.

\section{Modeling}
\label{sec:modeling}
We need a model that connects attitude commands to forces, forces
to relative motion, and relative motion to separation. The satellites in formation each occupy a ``slot,'' with each slot specifying a relative orbit. Attitude changes the projected areas and directions of
aerodynamic lift, drag, and solar radiation pressure (SRP). Common
forces move the cluster; their differential components
change its shape. At mean solar density, broadside drag produces $8.5\times10^{-7}$~m/s$^2$, comparable to the illuminated SRP scale of $9.1\times10^{-7}$~m/s$^2$. Control authority is dependent on how much these forces change over the permissible attitude range. Its effect varies with orbital phase and differs by element: longitude
is much easier to change than inclination. 

\begin{table}[tb]\centering\small
\caption{Vehicle and orbit.}
\label{tab:vehicle}
\begin{tabular}{lr}
\toprule
array, two wings$^{a}$ & $2\times16.75\times3$~m $=100.5$~m$^2$ \\
broadside, with bus$^{a}$ & $105$~m$^2$ \\
bus$^{b}$ & $3\times1.5\times0.25$~m \\
tip-to-tip span$^{b}$ & $\approx35$~m \\
dry mass$^{a}$ & $575$~kg \\
broadside $C_D$, finite-$S$$^{c}$ & $2.335$ \\
ballistic coefficient$^{c}$ & $\approx2.35$~kg/m$^2$ \\
\midrule
altitude$^{d}$ & $650$~km \\
inclination$^{d}$ & $98^\circ$, SSO \\
local time$^{d}$ & dawn--dusk \\
lattice spacing$^{d}$ & $100$~m, $81$ slots in a disc \\
fleet size$^{d}$ & $81$ \\
\bottomrule
\end{tabular}
\vspace{2pt}
{\footnotesize\raggedright
$^{a}$Representative of a
Starlink~V2-class satellite. The results depend on ballistic coefficient and broadside-to-edge-on area ratio; the specific dimensions are a representative model, not a published configuration. The simulated model uses the values in this table. 
$^{b}$Assumed, not public.
$^{c}$Derived from geometry and force model.
$^{d}$From the concept of Ref.~\cite{concept}.\par}
\end{table}

\subsection{Modeling assumptions and limitations}
\label{sec:assumptions}

The results use the following assumptions. Numerical settings and campaign-specific inputs are given in Sec.~\ref{sec:campaigns}.
\begin{itemize}
\item \textbf{Dynamics.} The main nonlinear simulations propagate Cartesian position and velocity under central gravity, \(J_2\), aerodynamic forces, and SRP. Guidance uses linear mean-ROE dynamics, with the state conversion described in Sec.~\ref{sec:relstate}. Sensitivity to additional gravity terms is examined in Sec.~\ref{sec:omitted}. 
\item \textbf{Environment.} Solar level and Sun angle $\beta$ are fixed within each run and vary between cases. The Sun direction precesses with the nominal orbital plane to maintain constant $\beta$. No mission epoch is prescribed. Density is from \cite{smad}. The nonlinear force model uses temperatures from 700 to 1100 K, but the controller’s aerodynamic predictor uses 900 K throughout.
\item \textbf{Mass.} Spacecraft mass is constant. For the isotropic grid, three years of
worst-satellite inclination maintenance and solar-maximum
reboost expends about $4$~kg, under one percent of vehicle
mass.
\item \textbf{Thrust realization.} In simulation, the inclination corrections are modeled as ideal orbit-normal impulses. Propellant estimates exclude the additional impulse that mounting, operational, or attitude constraints could require. For example, a Sun-aligned thrust axis would place a fraction $\cos\beta$ of the total impulse in-plane (26\% at $\beta=75^\circ$). This component is not carried in the results and the excess fuel is not included in $\Delta V$ estimates, as the actual implementation (thruster pairs, in-plane thrust, slewing to thrust, etc.) may vary.
\item \textbf{Navigation and links.} Centroid consensus estimation and closed-loop control are tested separately. Link topology and
terminal tracking rates are assumed. An integrated flight estimator fusing GNSS, orbital dynamics, and
optical-link geometry is not demonstrated
(Secs.~\ref{sec:control} and~\ref{sec:future}).
\item \textbf{Validation scope.} Closed-loop results cover the sampled conditions and up to 600 orbits. Global closed-loop stability has not been proved, and a mission-length
propagation remains open (Secs.~\ref{sec:stability}
and~\ref{sec:future}).
\end{itemize}

\subsection{Mean relative orbital elements}
\label{sec:relstate}

We describe relative motion with six orbital elements: relative semi-major axis and mean longitude, and two components each for relative eccentricity and inclination. Let $(a_k,e_k,I_k,\Omega_k,\omega_k,M_k)$ be the mean
orbital elements of vehicle $k$, with subscript $c$ denoting the virtual chief at the fleet centroid (not a specific spacecraft). Define
$u_k=M_k+\omega_k$ and
$\Delta q_k=q_k-q_c$ for a vehicle-minus-chief difference. 
The quasi-nonsingular
relative orbital elements (ROEs) are~\cite{koenig}
\begin{equation}
\delta\boldsymbol\alpha_k=
\begin{bmatrix}
\delta a_k\\ \delta\lambda_k\\
\delta e_{x,k}\\ \delta e_{y,k}\\
\delta i_{x,k}\\ \delta i_{y,k}
\end{bmatrix}
=\begin{bmatrix}
\Delta a_k/a_c\\
\Delta u_k+\Delta\Omega_k\cos I_c\\
\Delta(e\cos\omega)_k\\
\Delta(e\sin\omega)_k\\
\Delta I_k\\
\Delta\Omega_k\sin I_c
\end{bmatrix}.
\label{eq:roe}
\end{equation}
The relative eccentricity and inclination vectors ($\delta \mathbf{e}$ and $\delta \mathbf{i}$) describe the in-plane relative geometry and the out-of-plane differences, respectively. The relative inclination vector contains inclination
difference in $x$ and node difference scaled by $\sin I_c$
in $y$. At zero eccentricity, the orbital phase $u$ is used directly because $M$ and $\omega$ are not individually defined. The combined angular difference in $\delta\lambda$ is wrapped to $\left[-\pi,\pi\right)$.

Numerically, ROEs are first computed relative to the member nearest the centroid, then recentered by subtracting their fleet mean. This gives the virtual-chief state to first order and is the state used for feedback. A check over all reference-member choices gave a maximum meter-scaled component variation of 0.036 m, in relative longitude. Distributed centroid estimation is described in Sec.~\ref{sec:control}.

Within the controller, coefficients are evaluated at the fixed nominal orbit $a=7028.14$ km, $e=0$, and $I=98^\circ$, with $n=1.0715\times10^{-3}$ rad/s and a period of 97.73 min. These model parameters remain fixed within each run while the simulated orbits evolve. The ROEs are dimensionless. The numerical conversion uses the common member’s semi-major axis to obtain meter-scaled coordinates; the analytical model uses the nominal value $a$. ROE magnitudes quoted in meters refer to these scaled coordinates.

The controller uses mean ROEs because the short-period $J_2$ oscillations in osculating elements would otherwise appear as errors to correct. We remove these terms using Lara’s first-order Brouwer short-period corrections in polar-nodal variables.\cite{lara2015} Both this correction and fleet-mean subtraction are applied to the measured state, designed slots, and propagated reference. The long-period corrections are not applied. The nonlinear simulations retain the short-period motion; the conversion provides the state used by guidance. Consistent slot initialization also matters: an unintended mean semi-major-axis offset produces secular along-track drift.

\subsection{ROE evolution}
\label{sec:evolution}

The relative orbits evolve under gravity and differential forces. We separate these contributions in the linear mean-element model as
\begin{equation}
\frac{\mathrm d\,\delta\boldsymbol\alpha_k}{\mathrm d\,u}
=\frac{1}{n}\left(A\,\delta\boldsymbol\alpha_k+B(u)\,\Delta\mathbf a_k\right). \label{eq:roe_dynamics}
\end{equation}
Guidance uses \(du/dt=n\), so the factor \(1/n\) converts the time rates defined below to rates per unit orbital phase. The matrix $A$ contains Keplerian and first-order secular $J_2$ drift.\cite{koenig}  The acceleration $\Delta\mathbf a_k$ is the vehicle’s nongravitational RTN acceleration minus the fleet mean, and $B(u)$ maps the acceleration to ROE rates.

Omitting vehicle subscripts, the time-domain unforced equations at the nominal orbit are
\begin{equation}\label{eq:drift}
\begin{aligned}
\dot{\delta a}&=0,\qquad \dot{\delta i_x}=0,\\
\dot{\delta\lambda}
&=-1.6026\times10^{-3}\delta a
+1.3818\times10^{-6}\delta i_x,\\
\dot{\delta e_x}&=+6.4679\times10^{-7}\delta e_y,\\
\dot{\delta e_y}&=-6.4679\times10^{-7}\delta e_x,\\
\dot{\delta i_y}
&=-6.9089\times10^{-7}\delta a
+1.4046\times10^{-6}\delta i_x.
\end{aligned}
\end{equation}
\noindent These are the six nonzero entries of $A$, in s$^{-1}$.
Semi-major-axis and inclination differences drive
\emph{longitude drift}; $\delta\mathbf e$ undergoes
\emph{apsidal rotation}; and $\delta\mathbf i$ undergoes
\emph{inclination shear}, with $\delta i_x$ fixed and
$\delta i_y$ evolving.

To leading order about a circular orbit,
\[
B(u)=\frac{1}{na}
\begin{bmatrix}
0&2&0\\ -2&0&0\\
\sin u&2\cos u&0\\ -\cos u&2\sin u&0\\
0&0&\cos u\\ 0&0&\sin u
\end{bmatrix},
\]
with columns ordered radial, along-track, and normal. Thus radial and along-track
forces affect the in-plane elements, while normal force
acts on inclination.

The homogeneous drift can be propagated analytically. Let $A_{\rm rot}$ contain the two eccentricity-rotation entries and $N=A-A_{\rm rot}$. Since $N^2=0$ and $A_{\rm rot}$ commutes with $N$, 
\begin{equation} \label{eq:phi}
\Phi(h)=e^{Ah}=e^{A_{\rm rot}h}(I+Nh),
\end{equation}
where $h=\Delta u/n$ is the interval duration in seconds. This is exact for the retained linear model, so guidance
needs no numerical integration of the drift. Including differential acceleration gives{\small
\begin{align}
\delta\boldsymbol\alpha_k(t+h)={}&
\Phi(h)\delta\boldsymbol\alpha_k(t)\nonumber\\
&+\int_0^h\Phi(h-\tau)B(u(t+\tau))
\Delta\mathbf a_k(t+\tau)\,\mathrm d\tau.
\label{eq:roe_flow}
\end{align}}
The attitude allocator uses midpoint quadrature for this
forcing integral (Sec.~\ref{sec:control}).

For closed-loop initialization, choose the initial semi-major-axis offsets so that their contribution to secular longitude drift cancels the contribution from inclination differences via
\begin{equation}
    a\,\delta a_k=-\frac{\partial \dot{\delta \lambda}/\partial\delta i_x}{\partial\dot{\delta\lambda}/\partial\delta a}\,a\,\delta i_{x,k}\simeq8.62\times10^{-4}\,a\,\delta i_{x,k}. \label{eq:insertion}
\end{equation}
\noindent This condition gives the linear longitude-drift balance. In the numerical initializer, we measure the secular longitude drift of the constructed trajectory and apply a tangential velocity correction, repeated for three iterations with re-evaluation. These corrections also slightly perturb the $\mathbf{\delta e}$ pattern. The refinement matters because a residual $1$~mm mean semi-major-axis error produces about $0.94$ m of longitude displacement over 100 orbits. These offsets are part of the desired formation, not tracking errors to remove. The remaining inclination shear motivates the choice of slot evolution in the next subsection.

\subsection{Desired slots and tracking error}
\label{sec:sloterr}

A \emph{slot} is a desired relative-orbit trajectory
$\delta\boldsymbol\alpha_k^{\rm des}(u)$. Its initial elements define the arrangement, and a reference law
specifies how each slot evolves. In the linear model,
\[
\frac{\rm d\delta\boldsymbol\alpha_k^{\rm des}}{\rm d u}
 =\frac{1}{n}A_{\rm ref}\delta\boldsymbol\alpha_k^{\rm des},
\qquad
\delta\boldsymbol\alpha_k^{\rm des}(u_0)
 =\delta\boldsymbol\alpha_{k,0}^{\rm des}.
\]
The matrix $A_{\rm ref}$ has the same time units as $A$.

We define the ROE tracking error as the meter-scaled difference between the vehicle's ROEs and its desired slot,
\begin{equation}
    x_k = a\left(\delta\boldsymbol\alpha_k-\delta\boldsymbol\alpha_k^{\rm des}\right),
\end{equation}
using the fixed nominal length scale $a$. With vehicle indices omitted, the components are
$x_a,x_\lambda,x_{e_x},x_{e_y},x_{i_x},x_{i_y}$,
with $\mathbf{x}_e=(x_{e_x},x_{e_y})^\top$. Subtracting the reference evolution from
Eq.~\ref{eq:roe_dynamics} gives
\begin{equation}
\begin{aligned}
\frac{dx_k}{du}
&=\frac{1}{n}\left[
Ax_k+a f_{{\rm slot},k}+aB(u)\Delta\mathbf a_k
\right],\\
f_{{\rm slot},k}
&=(A-A_{\rm ref})\delta\boldsymbol\alpha_k^{\rm des}.
\end{aligned}
\label{eq:sloterr}
\end{equation}
The term $f_{{\rm slot},k}$ is the difference between natural drift and desired slot motion. If it is nonzero, control is needed even when the satellite is exactly on its slot. Choosing $A_{\rm ref}=A$ removes this reference demand but allows the inclination pattern to shear. Freezing the reference with $A_{\rm ref}=0$ requires control to oppose all natural drift. A frozen reference, tested at $\beta=75^\circ$ and mean solar level, left $53.9$~m RMS slot error after $110$ orbits
and $3742$ burns. We use co-rotation, $A_{\rm ref}=A_{\rm cr}$, because pairwise minimum separation depends on the angle between $\Delta\delta\mathbf{e}$ and $\Delta\delta\mathbf{i}$, which apsidal rotation opens (Sec.~\ref{sec:refsafety}).
Table~\ref{tab:acr_definition} defines $A_{\rm cr}$ by replacing four
coefficients in $A$; every other entry is unchanged. Here
$\dot\omega=-6.4679\times10^{-7}$~s$^{-1}$ is the $\mathbf{\delta e}$
rotation rate. The factor $8.62\times10^{-4}$ is the rounded longitude-drift
matching ratio from Sec.~\ref{sec:evolution}.

\begin{table}[tb]
\centering\footnotesize
\setlength{\tabcolsep}{3pt}
\caption{Definition of $A_{\rm cr}$ from $A$. Each entry is the coefficient
multiplying the indicated state term in the indicated time-rate equation.
Coefficients are in s$^{-1}$ and numerical values are rounded. All unlisted
entries are unchanged.}
\label{tab:acr_definition}
\begin{tabular}{@{}llll@{}}
\toprule
Rate & State term & $A$ & $A_{\rm cr}$ \\
\midrule
$\dot{\delta i_x}$ & $\delta i_y$ & $0$ & $-\dot\omega$ \\
$\dot{\delta i_y}$ & $\delta i_x$ & $1.4046\times10^{-6}$ & $\dot\omega$ \\
$\dot{\delta i_y}$ & $\delta a$ & $-6.9089\times10^{-7}$ & $0$ \\
$\dot{\delta a}$ & $\delta i_y$ & $0$ & $-(8.62\times10^{-4})\dot\omega$ \\
\bottomrule
\end{tabular}
\end{table}

The first three changes impose inclination rotation and remove the natural
semi-major-axis contribution to nodal drift from the reference. The fourth
makes the desired semi-major axis follow the rotating inclination pattern,
preserving the initial linear longitude-drift balance. The eccentricity and
longitude rows remain unchanged. Secs.~\ref{sec:refsafety} and ~\ref{sec:dv} examine the resulting separation margin and inclination-maintenance cost.

\subsection{Forces and actuation}
We model aerodynamic forces with a free-molecular flat-plate
model at finite speed ratio~\cite{storch} and solar radiation pressure
with a specular/diffuse plate. A cylindrical
Earth-shadow model excludes SRP during eclipse. We assume an atmosphere corotating with Earth. Appendix~\ref{app:forces} gives the relative-wind geometry and force-model coefficients. 

The baseline attitude is Sun-pointing,\footnote{
We tested a bias away from the drag minimum at near-polar Sun,
but it worsened performance there; the offset is therefore
zero throughout.} and attitude commands tilt the spacecraft away from this orientation to modulate aerodynamic and SRP forces for in-plane control. The maximum commanded off-pointing is $\theta_{\rm max}=20^\circ$, limiting the solar arrays'
projected-area cosine loss to approximately $6\%$.

For Sun angle $\beta$ above the orbital plane, the Sun-pointing array normal in RTN is
\begin{equation}
    \hat n_{\rm ref}(u) = (\cos\beta\cos u,\ -\cos\beta\sin u,\ \sin\beta).
\end{equation}
\noindent We choose the Sun phase so that its in-plane projection points radially outward at $u=0$ (local solar noon). 

The baseline pitch, measured from the radial--normal (RN) plane toward the along-track direction, tracks the sun and sweeps $\pm\left(90^\circ-\beta\right)$ over each orbit. The command $c=(c_1,c_2)^\top$, with $\|c\|\le1$,
specifies a tilt from the Sun-pointing baseline, with
$c_1$ and $c_2$ along the local pitch and roll directions,
respectively.
Let $\hat T$ be the along-track unit vector and define
the local roll and pitch directions as
\[
\hat e_2=
\frac{\hat T\times\hat n_{\rm ref}}
     {\|\hat T\times\hat n_{\rm ref}\|},
\qquad
\hat e_1=\hat n_{\rm ref}\times\hat e_2.
\]
\noindent For $\|c\|>0$, the commanded normal is
{\small
\[
\hat n(u,c)=
\hat n_{\rm ref}(u)\cos(\theta_{\rm max}\|c\|)
+\frac{c_1\hat e_1+c_2\hat e_2}{\|c\|}\sin(\theta_{\rm max} \|c\|),
\]
}
\noindent
and
$\hat n(u,0)=\hat n_{\rm ref}(u)$.
The command direction selects the combination of pitch
and roll, while its magnitude sets the off-pointing
angle $\theta_{\rm max} \|c\|$. Rotation about the array normal (yaw) is not commanded.
The ideal plate forces are independent of yaw, while
the bus contribution to drag depends on yaw.\footnote{Measured at mean solar activity and $\beta=75^\circ$ for a vehicle with full facets, around a full $2\pi$ rotation, the along-track drag varies from 1.05\% peak-to-peak at $u=0$ up to $4.6$\% when the flat-plate approximation is already in drag-minimum configuration. The orbit-mean spread is 1.1\% of the vehicle's own along-track drag. The control laws use a flat-plate approximation and yaw has no effect.}

To predict the effect of a command, we compare its
aerodynamic and SRP acceleration with that of the same
vehicle remaining Sun-pointing. Let $\mathbf a_{\rm RTN}(u,c)$ denote the modeled
aerodynamic and SRP acceleration at orbital phase $u$
and command $c$. The corresponding change in mean-ROE rate is
\begin{equation}\label{eq:bnl}
b_{nl}(u,c)=B(u)\left(
\mathbf a_{\rm RTN}(u,c)-\mathbf a_{\rm RTN}(u,0)
\right),
\end{equation}
\noindent with units of $s^{-1}$ and $b_{nl}(u,0)=0$. This predicts only the individual vehicle's own response $b_{nl}$, although the measured ROEs are fleet-centered. No disturbance estimate or
scheduled thrust input is included in the attitude
prediction. Model mismatch and impulses affect subsequent
state measurements and are dealt with via the control law feedback (Sec.~\ref{sec:control}).

To quantify the acceleration available from attitude changes, we sampled 48 command directions at the full $20^\circ$
tilt over one orbit. At $\beta=75^\circ$, the largest
acceleration change relative to Sun-pointing increases
from $1.05\times10^{-7}$~m/s$^2$ at solar minimum to
$1.09\times10^{-6}$~m/s$^2$ at solar maximum. However, the commands do not provide equal authority in both
directions. At $\beta=90^\circ$, the Sun-pointing arrays
are nearly edge-on to the relative wind, leaving more
room to increase drag than to reduce it. At mean solar
activity, the largest sampled drag increase is about
13 times the largest decrease. The normal response
can also be one-sided: at $\beta=75^\circ$ and solar
minimum, all sampled full-tilt commands produce a
positive normal acceleration change.

We use scheduled orbit-normal impulses for inclination
maintenance. The assumed thruster specifications are
based on a flown argon Hall thruster: $170$~mN thrust,
$2500$~s specific impulse, $4.2$~kW power, and $2.1$~kg
mass.\cite{spacex-thruster}

\subsection{Candidate geometries}

\begin{figure*}[tb]\centering
\includegraphics[width=0.85\textwidth]{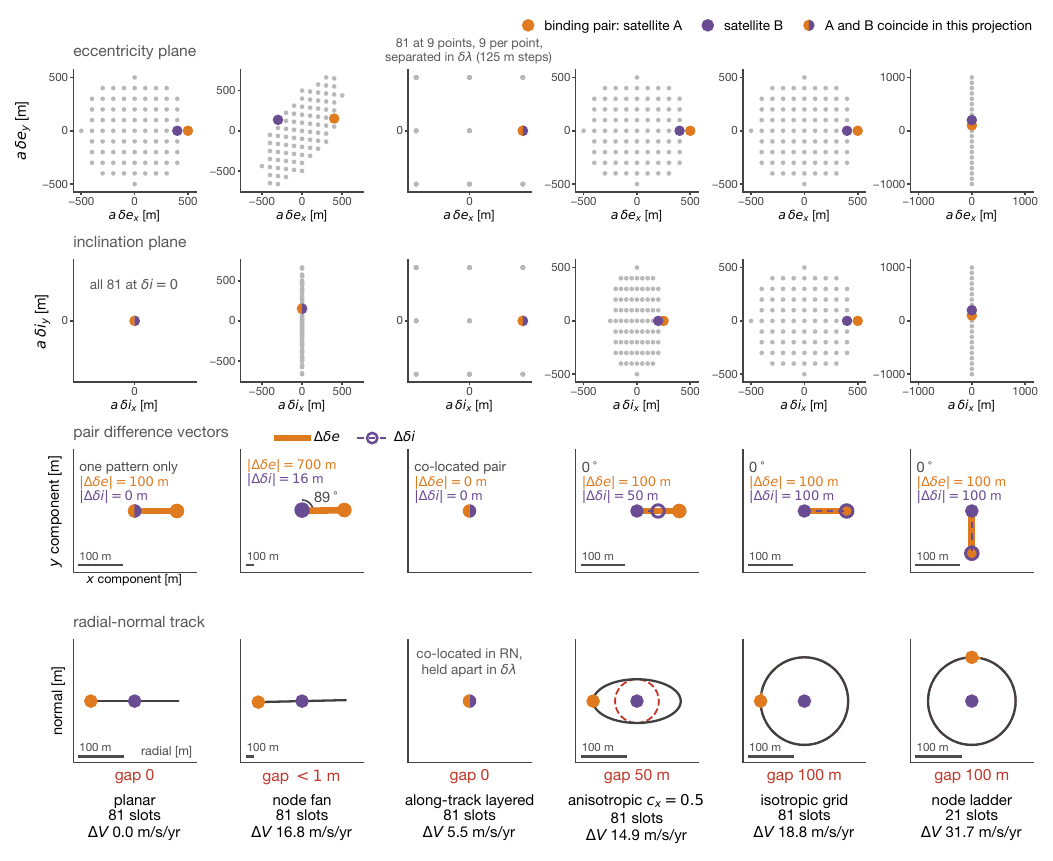}
\caption{Candidate arrangements. Top two rows: all satellites in the eccentricity and inclination planes, with the binding pair highlighted in orange and purple; a half-and-half disc marks a projection in which the two coincide. Third row: that pair's difference vectors
and the angle between them. Fourth row: its RN
trajectory and minimum separation, which the co-rotating
reference holds at every apsidal phase. Capacity and
measured annual inclination cost, $\beta=75^\circ$ and mid solar level, beneath each design. An additional tested case, omitted from the figure, is the anisotropic grid with $c_x=0.7$.}
\label{fig:arrangements}\end{figure*}

The slot geometry determines how satellites
are separated in longitude and in the RN
plane. Radial--normal protection depends on the magnitudes
and relative orientation of each pair's $\mathbf{\delta e}$ and
$\mathbf{\delta i}$ vectors, following the
eccentricity--inclination separation principle of
Ref.~\cite{damico}. We compare seven arrangements with different
$\mathbf{\delta e}$ and $\mathbf{\delta i}$ patterns. Fig.~\ref{fig:arrangements}
shows six of the arrangements, together with their
minimum gaps, capacities, and inclination-maintenance
costs.

Except for the layered and ladder cases, the designs start
from $(a\delta e_x,a\delta e_y)=100(p,q)$~m, with integer
$p,q$ satisfying $p^2+q^2\le25$. This gives 81 points
within a $500$~m radius. Table~\ref{tab:geom_recipes}
specifies the designed geometries, with
$\delta a=\delta\lambda=0$ unless stated otherwise.
Closed-loop initialization corrections are described
in Sec.~\ref{sec:evolution}.

\begin{table}[tb]
\centering\small
\setlength{\tabcolsep}{3pt}
\renewcommand{\arraystretch}{1.1}
\caption{Initial slot constructions. The grid-family row
includes the two anisotropic cases and the isotropic case.
The node-fan shear factor is $s_{\rm f}=0.88$.}
\label{tab:geom_recipes}
\begin{tabular}{@{}p{0.27\columnwidth}p{0.67\columnwidth}@{}}
\toprule
Arrangement & Construction \\
\midrule
Planar &
$\mathbf{\delta i}=0$. \\

Node fan &
Shear $\delta e_y\gets \delta e_y+s_{\rm f} \delta e_x$;
then $\delta i_x=0$, $\delta i_y=\delta e_y$. \\

Along-track layered &
A $3\times3$ grid with
$a\delta e_x, a\delta e_y\in\{-100,0,100\}$~m
and $\mathbf{\delta i}=\mathbf{\delta e}$.
At each point, nine layers:
$a\delta \lambda=125\ell$~m,
$\ell=-4,\ldots,4$. \\

Anisotropic/ isotropic grids &
$\delta i_x=c_x\delta e_x$, $\delta i_y=\delta e_y$,
with $c_x=0.5,\ 0.7,\ 1$.
The last is isotropic. \\

Node ladder &
$a\delta e_y=a\delta i_y=100k$~m,
$k=-10,\ldots,10$;
$\delta e_x=\delta i_x=0$. \\
\bottomrule
\end{tabular}
\end{table}

Planar pairs cross zero RN separation twice
per orbit and rely on along-track phasing. Layering puts
nine satellites on each RN track; it increases
capacity without increasing inclination spread, but their
separation lies along the direction of drift.

The node fan differs from the $c_x=0$ limit by shearing
the eccentricity grid before assigning node offsets.
No rescaling follows, so its $y$-extent grows from
$\pm500$ to $\pm664$~m. Both start without inclination
shear, but the fan has greater inclination spread and
still leaves some pairs with small gaps. Within the
grid family, increasing $c_x$ adds inclination spread
in $\delta i_x$, trading maintenance cost for
RN separation (Sec.~\ref{sec:trade}).

The node ladder provides pairwise RN separation
along a line, with $21$ slots over $\pm1$~km. Its initial
$\delta i_x=0$ eliminates inclination shear, but co-rotation
introduces $\delta i_x$ and reaches the inclination-ladder
orientation after a quarter cycle ($28.1$ days at the nominal
orbit). Both orientations have the same annualized cost
when averaged over an apsidal cycle.

\subsection{Simulation campaigns}
\label{sec:campaigns}

Table~\ref{tab:campaigns} separates three kinds of test.
Internal-model tests regulate one satellite to the origin without
navigation or pointing errors. Nonlinear fleet tests exercise the
controller and thrust schedule with $81$ satellites, a $1$~m navigation
bias, and the $3^\circ$ pointing distribution defined below.
Passive-fault trials measure separation after control loss, with
healthy vehicles maintained by the prescription below.

\begin{table}[tb]
\centering\scriptsize
\setlength{\tabcolsep}{3pt}
\caption{Simulation campaigns. Fleet tests use a co-rotating reference
except for the in-plane capture tests.
Lengths are in orbits unless days are specified.}
\label{tab:campaigns}
\begin{tabular}{@{}lrr@{}}
\toprule
Test & Runs & Length \\
\midrule
\multicolumn{3}{@{}l}{\emph{Internal model: 332 cases}} \\
Nominal acquisition & 36 & 40--90 \\
Envelope acquisition & 144 & 120--400 \\
Maintenance & 152 & 12 \\
\midrule
\multicolumn{3}{@{}l}{\emph{Nonlinear fleet}} \\
Nine-cell environment grid & 9 & 110 \\
Extended maintenance & 1 & 600 \\
$c_x$ sweep & 6 & 600 \\
Split rotation & 3 & 110 \\
$f_{\rm slot}$ ablation & 2 & 110 \\
Integration refinement & 3 & 110 \\
Capture & 4 & 400 \\
\midrule
Baseline passive-fault trials & 2000/leg & 30 d \\
\bottomrule
\end{tabular}
\end{table}

The nine-cell grid crosses $\beta=58.6^\circ,75^\circ,90^\circ$
with solar levels $0,0.5,1$. At $650$~km these levels correspond to
densities $1.32\times10^{-14}$, $6.87\times10^{-14}$, and
$3.08\times10^{-13}$~kg/m$^3$~\cite{smad}, and temperatures
$700$, $900$, and $1100$~K. Between these points, density is
interpolated logarithmically and temperature linearly. The plant
evaluates density at each vehicle's altitude; the controller uses
the nominal-altitude value and $900$~K.

Propagation uses fixed-step RK4: $128$ steps per orbit ($45.8$~s)
for fleet tests and $200$ ($29.3$~s) for passive trials. Fleet guidance
updates every $183.24$~s; reference tables use gravity-only propagation
with eight substeps per guidance interval. Nine-cell slot-error results are
fleet rms slot error at orbit $110$; extended-maintenance results use
$100$-orbit averages of the same quantity. At $\beta=75^\circ$ and mean solar level, reducing the fleet
step fourfold changed the window-mean error by less than $0.4\%$
and propellant by approximately $0.01\%$. The worst-satellite peak remained step-sensitive, increasing from $6.33$ to $8.37$~m.

Each passive leg fixes $\beta=75^\circ$ and $u=0$ orbital
phase. It starts from the untrimmed designed slots or, for release-epoch
tests, a maintained-fleet snapshot at orbit $0$, $200$, or $400$.
Solar level is drawn uniformly on $[0,1]$ once per run and held fixed;
this does not give a uniform density distribution. Each vehicle's
Sun-relative off-pointing magnitude is half-normal, with probability
$95\%$ at $\sigma=3^\circ$ and $5\%$ at $\sigma=30^\circ$, and
tilt direction uniform in the panel plane. These offsets remain fixed
relative to the prescribed Sun direction. Fleet control tests use
only the $3^\circ$ component. Single-fault modes choose the failed
vehicle uniformly. It retains its drawn attitude, receives an inertially fixed uniformly random attitude, or receives a $5$~cm/s impulse in an isotropic
direction, according to the fault mode.

In single-fault trials, healthy vehicles are reset each integration
step to a gravity-only reference table about the healthy fleet's
centroid, with independent Gaussian position offsets of $1$~m standard
deviation per axis, redrawn at every step. The table spans one orbit
at $200$ phases and is interpolated at the centroid's orbital phase.
The failed vehicle is excluded from that centroid and propagates
freely. Thus these trials assume prescribed healthy-fleet maintenance;
they do not run the controller or model temporally correlated
station-keeping errors. Under fleet-wide loss, all vehicles propagate
freely. Both $35$ and $70$~m keep-outs are tested using the analytic
minimum along each step's straight relative-position segment.

Separate dispersion comparisons use $1000$ trials per arm with matched
random draws. They add independent, uniform per-vehicle variations of
$\pm0.05$ in tangential accommodation and $\pm1\%$ each in mass and
area; these variations are absent from the  baseline safety campaigns.

\subsection{Sensitivity to omitted perturbations}
\label{sec:omitted}
We checked the effect of omitted gravity terms on the designed
$81$-satellite lattice using two uncontrolled, gravity-only propagations:
one with $J_2$, and one adding $J_3$--$J_6$ and lunisolar attraction.
The comparison used $320$ steps per orbit for $430$ orbits
($29.2$ days), with a static lunisolar ephemeris rather than a prescribed
mission epoch. For the closest designed pair, initially $100$~m apart,
the maximum-to-date relative-position discrepancy was $0.49$~m at
five days, $1.17$~m at ten days, and $9.79$~m at $29.2$ days.
The largest per-satellite residual about the fleet mean reached $49.76$~m at the final
horizon.\footnote{The cancellation is a scale law. A harmonic of degree $n$ has gradient length $r/n$, so a formation of size $L$ sees a differential of order $nL/r$ times the absolute displacement. For $L=1$~km across $J_3$-$J_6$, this is $4$--$9\times10^{-4}$, predicting $15$--$31$~m against the $49.8$~m measured. The perturbation effect therefore depends on formation scale. For this size, orbit, and horizon, the omitted terms mainly move the cluster as a whole.} The pair discrepancy is
$1.4\%$ of the $35$~m keep-out threshold at five days and $28\%$
at $29.2$ days. These are differences between propagated positions, rather than measured changes in minimum separation. The comparison gives a sensitivity scale; the safety Monte Carlo trials retain $J_2$ as their only gravity perturbation, and threshold-time changes were not evaluated.

\section{Control Architecture}
\label{sec:control}

We began this analysis anticipating formation-keeping to require
thrusters, with plume-impingement a concern at $100$~m. Attitude
modulation instead supplies most of the required relative-orbit
control through differential drag and SRP. Thrust remains for the inclination pattern and common reboost (Sec.~\ref{sec:dv}), which can avoid impingement. Reboost can be coordinated, and the
grids give $>60^\circ$ geometric clearance
from the assumed normal thrust axis.\footnote{The along-track
layered design is an exception: its longitude offsets
can cancel the along-track separation and place a
neighbor near the axis.}

\subsection{Objective}
At each update, the allocator (Sec.~\ref{ssec:alloc}) proposes attitude commands
and the controller selects the command with the lowest predicted
cost as defined here. Commands update $32$ times per orbit, with
$\Delta u=2\pi/32$ and $h=\Delta u/n=3.05$~min.
The selected attitude command is held throughout the interval.

The objective combines error and its tendency to grow. The same error can require different commands if the $A$ matrix (Eq.~\ref{eq:drift}) moves the satellite toward or away from its slot. The secular error
rates $v_\lambda,v_i$ are in m/day, from Eq.~\ref{eq:drift}. We combine these rates with their errors in a sliding-surface form as $s_\lambda=x_\lambda+K_\lambda v_\lambda$
and $s_i=x_{i_y}+K_i v_i$. Here $v_i$ is the secular rate of
$x_{i_y}-4.31\times10^{-4}x_\lambda$.
In $s_i$, we approximate this combination by $x_{i_y}$;
the omitted term is below $1$~mm RMS in the reported
maintained case. The resulting quadratic error metric is
{\small
\begin{equation}\label{eq:metric}
V(x)=\frac{1}{2}x^\top Px = \tfrac12\bigl[\,w_\lambda s_\lambda^{2}+w_i s_i^{2}
+w_e\|\mathbf{x}_e\|^{2}+\epsilon_i v_i^{2}\,\bigr].
\end{equation}
}
\noindent We use $K_\lambda=0.167$~d, $K_i=2.66$~d, $(w_\lambda,w_i,w_e)=(0.010,0.78,0.69)$~m$^{-2}$,
and $\epsilon_i=127$~(m/day)$^{-2}$.\footnote{
Capture radius $R$, priority, and patience $T$ give
$K=\sqrt{2\nu R/A_{\rm act}}$,
$w=\mathrm{priority}/R^2$, and
$\varepsilon=w\max(0,(R/(2\pi T A_{\rm act}))^2-K^2)$,
where $\nu$ is chain coupling and $A_{\rm act}$ is
worst-direction aerodynamic/SRP authority. These set
braking at $R$, position priority, and the cost of
rates persisting beyond $T$. The recipe
matches the implemented constants within $10$--$25\%$, with
implied horizons of $1.4$ and $3.9$ orbits; optimality and
parameter sensitivity remain open.} $V$ is positive
semidefinite. The longitude mode unpenalized by $V$ produces drift that is penalized at the predicted endpoint.

An attitude command $c$ is scored (Eq.~\ref{eq:metric}) by predicted
end-of-interval effect using midpoint quadrature. Using Eqs.~\ref{eq:phi}~and~\ref{eq:bnl},
\begin{equation}\label{eq:J}
\begin{aligned}
J(c)=\;&V\bigl(\Phi(h)\,x+h\,\Phi(h/2)\,a\,
   b_{nl}(u{+}\tfrac{\Delta u}{2},\,c)\bigr)\\
&+\gamma\|c\|^{2}
 +\tfrac12\,\delta_R\operatorname{tr}(H_\star)\,\|c\|^{2},
\end{aligned}
\end{equation}
where $\gamma$ penalizes large commands with little benefit. It decreases from $0.001$
at the knowledge radius $q$ (Sec.~\ref{sec:nav}, to avoid chasing navigation error) to zero at $4q$ and during capture.
The ridge $\operatorname{tr}(H_\star)$ conditions the candidate calculation below. The prediction omits $f_{\rm slot}$; including
it in a $\beta=75^\circ$, mean-solar case changed RMS
slot error from $3.04$ to $2.99$~m and propellant use
by $0.1\%$.

\subsection{Allocation and realization}\label{ssec:alloc}
A three-piece $C^1$ aerodynamic surrogate (App.~\ref{app:forces}, SRP remains exact) proposes commands. We evaluate every candidate with Eq.~\ref{eq:J} using the full nonlinear force model. This separates
the inexpensive search from the final comparison: surrogate
error can change the proposed commands, but all proposals
are judged by the same objective.

On each reachable piece $b$,
the allocator uses the quadratic approximation
$J_b(c)=q_b^\top c+\tfrac12c^\top R_bc+\mathrm{const}$,
with $G_b=ha\Phi(h/2)L_b$ and
$q_b=G_b^\top P\Phi(h)x$. Here $L_b$ is the linear
ROE-rate coefficient, so $G_bc$ is the corresponding
endpoint change. The vector $q_b$ gives the local slope
of the quadratic score; $R_b$ includes second-order
surrogate terms, command penalty, and ridge.
The candidates are:
\begin{itemize}
\item \emph{Coast:} $c=0$.
\item \emph{Piece KKT:} each reachable quadratic's
disk minimum, using feasible interior stationary points
$R_bc=-q_b$ and unit-circle stationary points with
$c=(\cos\theta,\sin\theta)$, obtained from the quartic
in $t=\tan(\theta/2)$ and $c=(-1,0)$, the point
omitted by finite-$t$.
\item \emph{Exact-tangent KKT:} the same disk solve
using the nonlinear model's derivative, with linear
term $g$ and Hessian $D_0^\top PD_0$
plus penalty and ridge.
\item \emph{Piece-boundary chords:} with incidence affine in $c$, each boundary is a chord of the command disc.
We check endpoints and any stationary point of
$J_b$ within the segment.
\item \emph{Cauchy:} $c_C=-rg/\|g\|$,
$r=\min(1,\|g\|/L)$. Initialize $L$ at $1.8$ times
the largest absolute active-piece Hessian eigenvalue;
double it until $J(c_C)\le J(0)-r\|g\|/2$.
Allow at most six doublings; use coast if the test still fails or when $g=0$.
\end{itemize}
Here $g=D_0^\top P\Phi(h)x$ and
$D_0=ha\Phi(h/2)\partial b_{nl}(u+\Delta u/2,c)/
\partial c|_{c=0}$.
We select $c^\star=\arg\min_{c\in\mathcal C}J(c)$.
Including coast ensures $J(c^\star)\le J(0)$.

A nearly flat command direction makes the quadratic
solve poorly conditioned. The ridge adds
$\delta_R\operatorname{tr}(H)I$ to each piece's
Hessian, while nonlinear scoring uses the common
$H_\star=D_0^\top PD_0$. Trace scaling keeps the
added curvature proportional to existing curvature.
We use $\delta_R=10^{-3}$ to condition the solve.

Pitch/roll rates are limited to $15^\circ$/min and
$5^\circ$/min. A two-control-step preview starts large,
opposite-tilt transitions early. Feedforward follows the moving Sun-pointing baseline; blending commands in the current attitude coordinates avoids treating that motion as a command jump.

\subsection{Navigation}
\label{sec:nav}

Control should stop correcting errors that navigation
cannot resolve. The attitude law compares the predicted RMS
position error over one orbit of free drift,
$s=\sqrt{x^\top M_{\rm pos}(u)x}$,
with the knowledge radius $q=1$~m, 
where $M_{\rm pos}=\tfrac1{2\pi}\int_0^{2\pi}
\Phi(\tau/n)^\top\allowbreak C(u+\tau)^\top\allowbreak C(u+\tau)
\Phi(\tau/n)\,\mathrm d\tau$.
The dimensionless $C$ has RTN rows
$(1,0,-\cos u,-\sin u,0,0)$,
$(0,1,2\sin u,-2\cos u,0,0)$, and
$(0,0,0,0,\sin u,-\cos u)$.
For $s<q$, attitude remains at baseline; outside,
feedback uses $(1-q/s)x$.

Using the fleet centroid as a
virtual chief avoids a single point
of failure and removes common errors, but the fleet must reach consensus.
Our separate consensus test reaches $1$~m in under
four minutes at $1$~Hz. With meter-scaled ROEs
$z_k$ about the nominal orbit, the centroid is $m=N_s^{-1}\sum_k z_k$. Vehicle $i$ uses its local centroid estimate $m_i$ to find its slot error $\widehat x_i=\widehat z_i-m_i-a\delta\boldsymbol\alpha_i^{\rm des}$. 

Graph averaging allows decentralized consensus~\cite{olfati}, with each satellite's estimate evolving by $m_i\leftarrow m_i+\epsilon\sum_{j\in \mathbb{N}_i}(m_j-m_i)$, where $\mathbb{N}_i$ is the set of vehicle $i$'s neighbors (1-4 per satellite). The averaging matrix 
$W=I-\epsilon L$ maps the estimates to their values after one exchange, with $\epsilon\simeq0.254$ and graph Laplacian $L$. The graph has $\lambda_2\simeq0.122$, giving a disagreement
decay factor $1-\epsilon\lambda_2\simeq0.969$ per exchange. Common error translates the lattice, but disagreement
distorts it, so we want navigation-limited convergence $\max_{i,j}\|m_i-m_j\|\ll q$.
Contraction by $0.969$ predicts $230$ rounds ($<4$ min at 1 Hz) until convergence from
$1414$~m, versus $227$ measured; $50\%$ link loss
requires $454$. This separate test is not a navigation
error budget. A flight estimator would also need GNSS and ISL
geometry fusion with outlier rejection.

\subsection{Inclination control}\label{sec:deadband}
Attitude has ample in-plane authority and little normal authority, so we treat the orbit-normal thrust control separately. An error exceeding a $3$~m deadband arms a normal burn at the
next node for $x_{i_x}$, or anti-node for
$x_{i_y}$, in $4.43$~mm/s ($15$~s) quanta.
The $q=1$~m knowledge radius projects to $1.4$~m
in $x_{i_y}$. Too narrow a band causes repeat
firing as it chatters and chases navigation error. Internal-model tests spent $131$ unnecessary quanta
at $2$~m and none at $3$~m, with smaller residual
and unchanged retention. The fleet campaigns use
$3$~m; improved crosslink-based relative positioning could permit $2$~m.

\subsection{Evidence and limitations}
\label{sec:stability}
\label{sec:evidence}

\begin{figure}[tb]\centering
\includegraphics[width=0.95\linewidth]{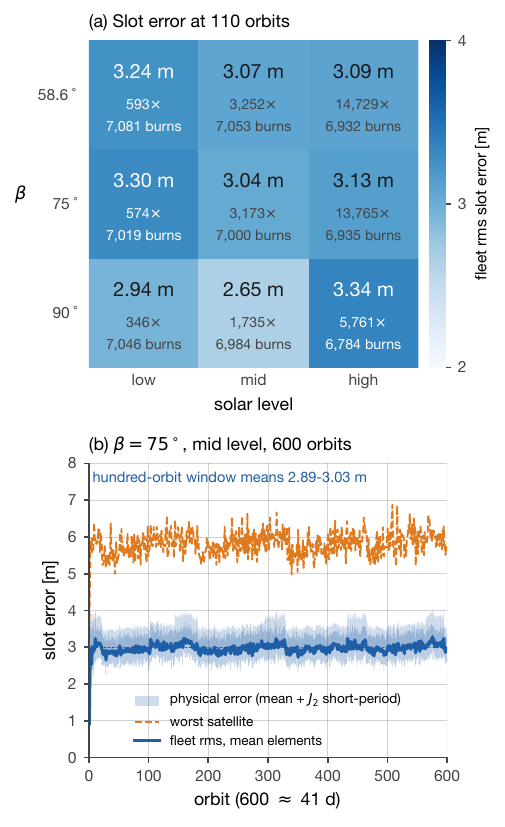}
\caption{(a) Closed-loop slot error at $110$ orbits across Sun
angle and solar level, $81$ satellites with a $3^\circ$
pointing dispersion, with open-to-closed rejection ratio and
burn count per cell. (b) The $\beta=75^\circ$, mid-solar-level case
over $600$ orbits: fleet rms and worst satellite.}
\label{fig:grid}\end{figure}

The internal-model tests check acquisition and retention under
the controller's assumed dynamics. We then test the law on the
nonlinear fleet model, where the force prediction and realized
attitude can differ. Internal-model tests pass all 332 cases, with and
without thrust. Thrust roughly halves acquisition time without
changing retention. At 110 orbits, slot error in nonlinear tests is $2.7$--$3.3$~m RMS: $1.2$~m
in-plane ROE error, with inclination held to its bands, rejecting
open-loop drift by $350$--$15{,}000$ times
(Fig.~\ref{fig:grid}).

The deployment study increases the initial scatter to test
acquisition on the nonlinear plant, beyond the small errors used in the internal-model tests.
Baseline capture succeeds from $26$~m RMS within
$400$ orbits, but not $35$~m. Longitude braking
matches authority only to $160$~m of longitude error when adding drag
and $11$~m when removing it; beyond these, late braking
causes ringing. Matching the asymmetric gain on the sliding surface doubles capture to
$52$~m. Scatter generates semi-major-axis error
$1.6$ times its per-axis value and longitude drift
of magnitude $3\pi |x_a|$ per orbit. 

An initial fleet-relative
correction $|\Delta v_T|=n|x_a|/2$ captures from
$100$~m/axis ($173$~m RMS) for a $\Delta V$ cost of $8.7$~cm/s per
satellite, under $1\%$ of
annual inclination maintenance. An along-track
$1$~cm/s dispersion gives $19$~m of $x_a$. Several satellites crossed into others' keep-out during capture, so deployment would need path planning to avoid conjunction.

Over $600$ orbits ($41$ days), successive $100$-orbit
mean errors are $2.89,3.03,2.94,2.97,2.98,3.03$~m:
$3\%$ variation without secular growth. In every window, rim-pair
angle and minimum RN gap stay within $2^\circ$ and $3$~m of the reference.

We have not yet proved global stability of the full controller. Saturation prevents extending
the quadratic certificate over acquisition. The nonlinear tests provide evidence of acquisition and bounded tracking error over the sampled conditions. A separate local check, with attitude feedback active and thrust and dead-band scheduling disabled, gives a one-orbit linearized error map with spectral radius $0.5643$ and satisfies a discrete Lyapunov condition to machine precision. This supports local stability of the linearized attitude loop. It does not cover the knowledge gate, where the law stops correcting errors that navigation cannot resolve, or the saturated acquisition regime.

\section{Safety Characterization}
\label{sec:safety}
After control loss, how long does the formation remain
safe? The main fault campaigns start from the designed
lattice, with prescribed maintenance of healthy vehicles
(Sec.~\ref{sec:campaigns}). A separate uncontrolled-attitude test checks releases
from the maintained formation
(Sec.~\ref{sec:refsafety}).

\subsection{Metric and convention}

The coast time $T_{cf}$ ends at the first pairwise
keep-out entry. Survival is
$S(t)=\Pr(T_{cf}>t)$. We use $35$~m center-to-center 
for a $35$~m vehicle, and a $70$~m margin ring. An attitude-aware
boundary remains future work. Coasts use the dispersions in
Sec.~\ref{sec:campaigns} and are right-censored at $30$~days. We report $t_p=\sup\{t:S(t)\ge p\}$, where $t_{0.99}$ defines the response window
at $99\%$ survival, and $t_{0.90}$ compares designs. Fig.~\ref{fig:faultpanel} shows the survival curves and Table~\ref{tab:safety} gives the five-day values.

\subsection{Checking the separation margin}
\label{sec:refsafety}

\begin{figure*}[tb]\centering
\includegraphics[width=0.9\textwidth]{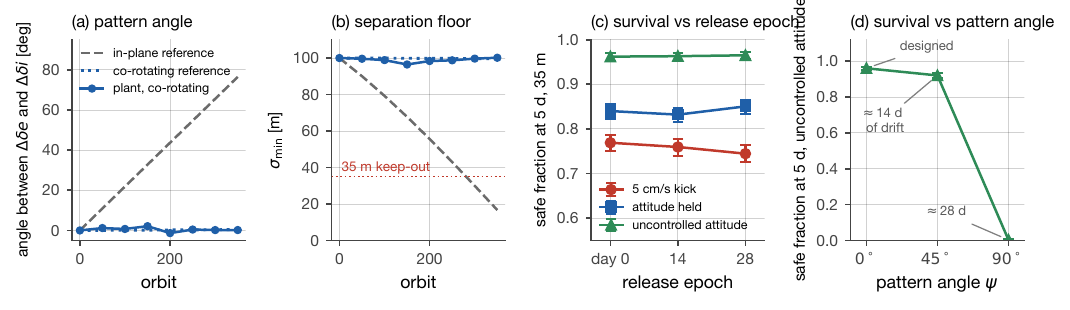}
\caption{Left pair: the adjacent rim pair over $350$ orbits
under the in-plane and co-rotating references, the angle
between $\Delta\delta e$ and $\Delta\delta i$ and the minimum
RN separation. Right pair: five-day survival of
the maintained lattice released at three epochs under the
co-rotating reference, and survival against the pattern
angle the in-plane reference would have reached.}
\label{fig:corotate}\end{figure*}

Tracking a slot does not by itself maintain separation. For pairwise
ROE differences, the leading-order minimum RN separation
over an orbit is
\[
d_{ij,\min}=\sigma_{\min}\!\left(a
\begin{bmatrix}
-\Delta \delta e_x & -\Delta \delta e_y\\
-\Delta \delta i_y & \Delta \delta i_x
\end{bmatrix}\right),
\]
where $\Delta$ denotes vehicle $i$ minus vehicle $j$, $\sigma_{\min}$
is the smallest singular value, and the small radial offset from the
semi-major-axis bias is neglected. A positive minimum requires
$\Delta\delta\mathbf e\cdot\Delta\delta\mathbf i\ne0$. Holding $\mathbf{\delta i}$ while $\mathbf{\delta e}$ rotates at $3.2^\circ$/day gives an initially aligned equal-amplitude pair
$d_{ij,\min}(t)=d\sqrt{1-|\sin(\dot\omega t)|}$.
The isotropic grid's $100$~m pair loses $35$~m
RN protection on days $19$--$37$ of each
$56$-day half-cycle, reaching zero near day $28$; an encounter also
requires along-track coincidence. Co-rotation preserves both vector lengths and their
angle, keeping its minimum RN separation unchanged
as the pattern rotates (Fig.~\ref{fig:corotate}a,b). We adopt it at $100$~m spacing, at about twice the inclination-maintenance cost of holding $\mathbf{\delta i}$ fixed
(Sec.~\ref{sec:dv}). Alternatives include periodic re-phasing or
phase-dependent detection and response. Doubling spacing to $200$~m
would reduce the unprotected interval from eighteen
to nine days per half-cycle. 

For the
$81$-satellite co-rotating grid, checking all $3240$ pairs every $50$ orbits (from $0$ to $350$) gives a
minimum RN separation of $93.9$~m. The largest reduction of any pair's reference separation was $8.7$~m. For the uncontrolled-attitude fault, five-day survival at the $35$~m keep-out was $0.962$, $0.963$, and $0.965$ at
the three maintained-state release epochs
(Fig.~\ref{fig:corotate}c).

\subsection{A single failed vehicle}
\label{sec:singlefault}

In a single failure, one satellite stops holding its slot
while the others follow the $1$~m model (details in Sec.~\ref{sec:campaigns}). 
We test three single-fault modes: a $5$~cm/s \emph{kick}
representing failure mid-maneuver; \emph{attitude held},
where the failed vehicle holds its Sun-pointing baseline (including its drawn pointing offset) without attitude-based correction; and \emph{uncontrolled
attitude}, using a random fixed orientation to represent
attitude-control loss. A held vehicle stays
nearer its neighbors while an uncontrolled one leaves faster.
The two no-impulse modes set the passive-geometry design
cases. The kick tests robustness
and likely requires active avoidance maneuvers, beyond the scope of this paper. Its survival improves with spread, which gives kick-induced excursions more room.

\begin{table}[tb]\centering\scriptsize
\caption{Fraction of $2000$ trials with no keep-out entry through
five days, by design and fault mode, at both keep-outs.
Common random numbers across cells; $95\%$ Wilson intervals within $\pm0.022$ at mid-range and $\pm0.005$ near $0$ or $1$.}
\label{tab:safety}
\setlength{\tabcolsep}{3pt}
\begin{tabular}{@{}lcccccccc@{}}
\toprule
 & \multicolumn{4}{c}{$35$~m keep-out} & \multicolumn{4}{c}{$70$~m keep-out} \\
\cmidrule(lr){2-5}\cmidrule(lr){6-9} & fleet & kick & held & unctl. & fleet & kick & held & unctl. \\
\midrule
planar & 0.000 & 0.002 & 0.000 & 0.000 & 0.000 & 0.000 & 0.000 & 0.000 \\
node fan & 0.000 & 0.246 & 0.008 & 0.046 & 0.000 & 0.021 & 0.000 & 0.001 \\
layered & 0.000 & 0.208 & 0.071 & 0.107 & 0.000 & 0.069 & 0.066 & 0.102 \\
anis.\ $0.5$ & 0.935 & 0.648 & 0.820 & 0.952 & 0.000 & 0.203 & 0.024 & 0.047 \\
anis.\ $0.7$ & 0.994 & 0.711 & 0.861 & 0.962 & 0.000 & 0.232 & 0.120 & 0.263 \\
iso.\ grid & 0.996 & 0.774 & 0.846 & 0.961 & 0.117 & 0.254 & 0.602 & 0.868 \\
node ladder & 1.000 & 0.833 & 0.973 & 0.990 & 0.999 & 0.392 & 0.707 & 0.877 \\
\bottomrule
\end{tabular}
\end{table}

\begin{figure*}[tb]\centering
\includegraphics[width=0.95\textwidth]{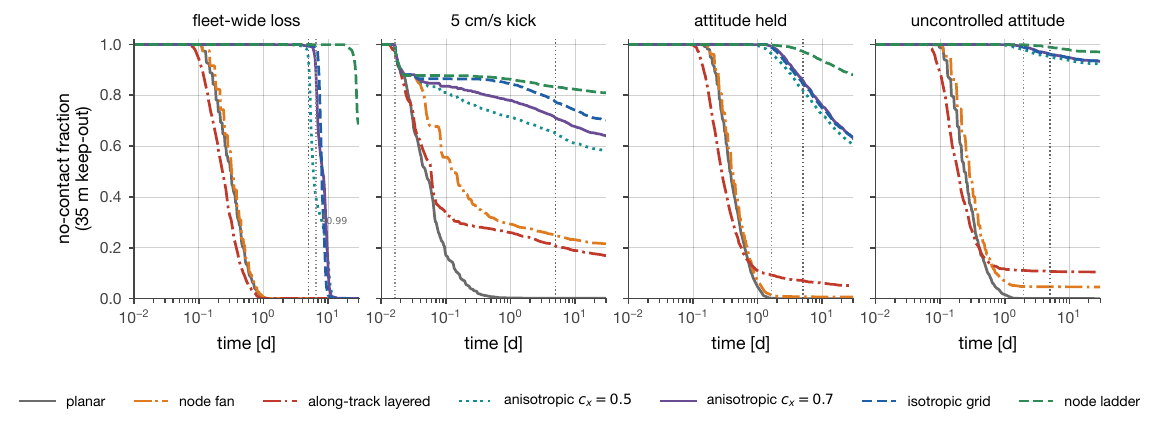}
\caption{Survival at the $35$~m keep-out against coast time, by configuration and fault mode, $2000$ trials per curve, $30$-day horizon, dotted line at five days. Left to right: fleet-wide control loss, a single vehicle with a $5$~cm/s kick, one holding attitude, one with uncontrolled fixed attitude; the dotted line at $t_{0.99}$ marks the isotropic grid's response window. The node ladder and the anisotropic grids survive an uncontrolled drifter; the along-track layered and planar designs contact within hours.}
\label{fig:faultpanel}\end{figure*}

Many curves (Fig.~\ref{fig:faultpanel}) show a ``knee'' where survival remains near unity for
an initial interval, then falls more rapidly as relative
motion consumes the available separation margin. The timing
and sharpness of this transition depend on the geometry
and fault mode. At five days and $35$~m keep-out, the ladder leads in every mode. Most isotropic/anisotropic trials remain clear, while most planar, node fan, and
layered do not. The shared-track layers rely on longitude separation, which a drifter traverses in hours. Rankings change at $70$~m: uncontrolled-attitude survival for $c_x=0.7$
falls from $0.962$ to $0.263$, versus $0.961$ to $0.868$
for the isotropic grid.

Along-track dispersion (increased with an uncontrolled attitude) sweeps drifters past their neighbors, improving safety in these trials. A kick's normal excursion $\delta v_N/n$ has median $23$~m and maximum $47$~m. Whether this reaches a neighbor or open space depends on the direction.

\subsection{Total control loss}
\label{sec:commonmode}

Under fleet-wide loss, the baseline grid shows a particularly
sharp knee. At the $35$~m keep-out, survival remains at $0.99$ through $6.49$~days, crosses $0.90$ at $7.20$ and $0.50$ at $7.94$. Half the distribution lies within a day and a half because the closest-pair envelope crosses the keep-out over a short interval.

Consider an initially aligned pair along
$\delta e_x$ and $\delta i_x$, with meter-scaled amplitude
$d=100$~m and no semi-major-axis difference. Applying the
secular dynamics of Sec.~\ref{sec:evolution} to the
separation expression in Sec.~\ref{sec:refsafety} gives
a first $35$~m RN crossing at $7.27$~days.
The simpler cosine approximation,
$d_{\min}(t)\simeq
d\cos\!\left(|\dot\omega|t+\arctan(\kappa_t t)\right)$,
with $\kappa_t=1.4046\times10^{-6}$~s$^{-1}$ the
inclination-shear coefficient, gives $7.96$~days.
Both estimates place the loss of RN protection
near the simulated fleet-loss median of $7.94$~days. Actual keep-out entry also requires along-track coincidence;
differential drag changes which pairs meet this condition.

Adding accommodation, mass, and area dispersions to the
$c_x=0.5/0.7$ fleet-loss cases shifts five-day survival
by $-0.009/+0.001$ and median coast time by
$+0.03/+0.30$ days. Every interval includes zero at
$1000$ trials per arm; no threshold shift was statistically resolved.

\subsection{Response windows}
\label{sec:latency}
Taking the isotropic grid at $99\%$ survival ($35$~m keep-out) as an
example, we find $1.6$~d of warning with attitude held, $1.9$~d with uncontrolled attitude, $6.5$~d under fleet loss, and 23 min after a kick. For this geometry, the held-attitude case sets the shorter
no-impulse response window. The kick result shows that detection alone is insufficient in such cases and active avoidance is required.

\section{Stationkeeping $\Delta V$ Budget}
\label{sec:dv}
For the isotropic grid, a three-year mean-solar scenario
requires approximately $78$~m/s per satellite at the fleet
mean and $108$~m/s at the rim, corresponding to about
$1.8$ and $2.5$~kg of propellant. The totals, shown in Table~\ref{tab:total_budget}, range from
$66$--$132$~m/s at the fleet mean and $96$--$162$~m/s
at the rim; the upper rim case requires about $3.8$~kg. They assume the ideal thrust-realization of Sec.~\ref{sec:assumptions} and magnetic (propellantless) desaturation of reaction wheels. Unless stated otherwise, annual $\Delta V$ costs
are fleet means per satellite.
\begin{table}[tb]
\centering\small
\setlength{\tabcolsep}{3pt}
\caption{Three-year budgets per satellite (m/s) at fixed solar
minimum/mean/maximum. Worst estimates scale the cycle-averaged
$29$~m/s/y at $500$~m with maximum inclination amplitude.
All totals include reboost of $3.42$/$15.73$/$69.43$
(including the $14\%$ allowance), $5.64$ for avoidance,
and $0.087$ for capture.}
\label{tab:total_budget}
\begin{tabular}{@{}lrrrrrr@{}}
\toprule
Design & \multicolumn{3}{c}{Fleet mean}
       & \multicolumn{3}{c}{Worst (est.)} \\
\cmidrule(lr){2-4}\cmidrule(l){5-7}
Solar & Min. & Mean & Max. & Min. & Mean & Max. \\
\midrule
Planar      & 9.1   & 21.5  & 75.2  & 9.1   & 21.5  & 75.2  \\
Node fan    & 59.5  & 71.9  & 125.6 & 124.7 & 137.0 & 190.7 \\
Layered     & 25.6  & 38.0  & 91.7  & 33.8  & 46.1  & 99.8  \\
$c_x=0.5$   & 53.8  & 66.2  & 119.9 & 96.1  & 108.5 & 162.2 \\
$c_x=0.7$   & 58.2  & 70.5  & 124.2 & 96.1  & 108.5 & 162.2 \\
Isotropic   & 65.5  & 77.9  & 131.6 & 96.1  & 108.5 & 162.2 \\
Node ladder & 104.2 & 116.6 & 170.3 & 183.1 & 195.5 & 249.2 \\
\bottomrule
\end{tabular}
\end{table}

\subsection{Inclination-maintenance cost}

Across nine conditions, inclination cost is $18.2$--$19.0$~m/s/y despite $23$-fold density variation and the full Sun-angle range. Geometry predicts shear cost alone at $10.21$~m/s/y (matching simulation to $5\%$). Rotating the $\mathbf{\delta i}$ pattern against it multiplies that by $1.92$ from the plant matrix, versus $1.93$ measured (18.8/9.7 m/s/yr). An ideal node/anti-node schedule averaged over the apsidal
cycle gives
\[
\Delta V_{\rm yr}=\frac{2}{\pi}\,n\,
(\kappa_t+2|\dot\omega|)\,a\overline{|\mathbf{\delta i}|}\,T_{\rm yr}
\simeq19.9~\mathrm{m/s},
\]
where $T_{\rm yr}$ is one year in seconds and
$\kappa_t,\dot\omega$ are defined in Secs.~\ref{sec:commonmode} and~\ref{sec:sloterr}. Agreement at roughly $5$--$6\%$
supports annualizing the $600$-orbit result. Propellant capacity must accommodate the worst-positioned (rim) satellite, which needs $29$~m/s per year averaged over the  cycle ($87$~m/s over three years, about $2$~kg of fuel), $1.5$ times the fleet mean, with a peak near $36$~m/s/y.

\subsection{Reboost}
Baseline-attitude (Sun-pointing) altitude maintenance, integrated over
the annual Sun-angle cycle, requires $1.0$, $4.6$, and
$20.3$~m/s/y at solar minimum, mean, and maximum per satellite. In the tested comparison, closed-loop attitude control increased
fleet decay by about $14\%$. This gives about $3.4$--$69$~m/s for a three-year life, depending on solar cycle, budgeted separately from
the simulated control burns. Polar Sun minimizes reboost because the
array is edge-on, but leaves attitude authority one-sided.

\subsection{Collision avoidance}\label{ssec:collision}

We reserve propellant for emergency avoidance of a failed member or external object. At $\beta=75^\circ$ and mean solar activity, the law's
held-command authority corresponds to
$|a_T|\simeq1.31\times10^{-7}$~m/s$^2$ in the
drag-increasing direction. The secular estimate
$a|\delta \lambda|\simeq\tfrac32|a_T|t^2$ gives
$6.7$~m after one orbit, $27.0$~m after two, and
$35$~m in $2.3$ orbits ($3.7$~h), without propellant.
The opposite direction is about fourteen times weaker
and takes approximately $14$~h for the same displacement. For shorter warning,
a normal impulse gives amplitude $a\left|\mathbf{\delta i}\right|=\Delta V/n$
after a quarter orbit: $35$~m requires $37.5$~mm/s,
a $127$~s burn. An assumed
avoidance reserve of $1.88$~m/s/y ($0.13$~kg over three
years) covers about $50$ such maneuvers per year.
Actual clearance depends on encounter phase and warning;
the hours-long attitude-response scale makes detection
time critical.

\section{Trade Study}
\label{sec:trade}

\begin{figure*}[tb]\centering
\includegraphics[width=0.85\textwidth]{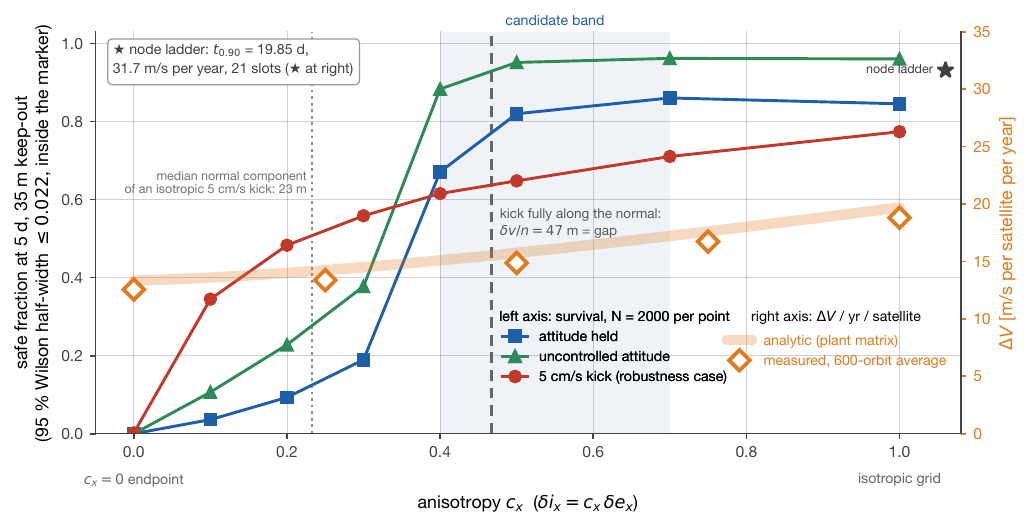}
\caption{Five-day survival at the $35$~m keep-out against
$c_x$ under three single-fault modes, $2000$ trials per point
on common random numbers, with measured co-rotating inclination $\Delta V$ per satellite per year ($600$-orbit averages, $\beta=75^\circ$, mid solar level) and its analytic prediction on the right axis. Shaded: the candidate band. The dashed vertical line marks where a kick entirely along the normal, $\delta v/n=47$~m, equals the minimum RN gap; the dotted line, at $23$~m, the median normal component. The node ladder sits off the dial.}
\label{fig:frontier}\end{figure*}

We can combine the response windows of Sec.~\ref{sec:safety} with the inclination cost of Sec.~\ref{sec:dv} to compare geometries. The useful design is the one that supplies the required failure detection/response time at an acceptable $\Delta V$ cost.

The assumed failure mode changes the
preferred arrangement. A
single failed vehicle passing neighbors that hold their
slots is more demanding than fleet-wide loss, which
preserves relative geometry longer. Designing only for
fleet-wide loss would allow $6.5$~days (in the isotropic case) for response
instead of the single-fault case's $1.6$. We therefore
base the trade on the single-fault arm.

For the grid family, we vary $c_x$ in
$\delta i_x=c_x\delta e_x$, from $c_x=0$, which
retains only $\delta i_y$, to the isotropic grid
at $c_x=1$. The dial was evaluated in closed loop at six settings, and tracking varies by under $0.5$~m across them while the propellant runs from $12.6$ to $18.8$~m/s/y.

\begin{table}[!hb]\centering\small
\caption{The anisotropy dial under the co-rotating reference at $\beta=75^\circ$, mid solar level: fleet RMS slot error at orbit $600$ and fleet-mean
inclination $\Delta V$ per satellite per year,
annualized from the same run, the initial rim shear-cancellation cost, and the $0.90$ survival quantile with attitude held at the $35$~m keep-out. Zero initial rim shear does not imply zero co-rotation cost.
}
\label{tab:frontier}
\setlength{\tabcolsep}{3pt}
\begin{tabular}{@{}lrrrrr@{}}
\toprule
design & $c_x$ & RMS error & $\Delta V$  & Rim shear & $t_{0.90}$\\
 & & [m] & \multicolumn{2}{c}{[m/s/y]\quad\;} & hold\\
\midrule
$c_x=0$ endpoint & $0.00$ & $2.6$ & $12.6$ & $0.00$  & $0.22$~d \\
anisotropic 25   & $0.25$ & $2.8$ & $13.4$ & $5.94$  & $0.25$~d \\
anisotropic 50   & $0.50$ & $2.7$ & $14.9$ & $11.88$ & $2.84$~d \\
anisotropic 70   & $0.70$ & $2.8$ & $16.3$ & $16.63$ & $3.68$~d \\
anisotropic 75   & $0.75$ & $2.7$ & $16.7$ & $17.82$ & $3.58$~d \\
isotropic grid   & $1.00$ & $2.8$ & $18.8$ & $23.76$ & $3.38$~d \\
node ladder      & ---    & $3.2$ & $31.7$ & ---     & $19.85$~d \\
\bottomrule
\end{tabular}
\end{table}

The eight-point sweep (Fig.~\ref{fig:frontier})
shows an S-shaped transition in five-day survival (on no-impulse cases) versus $c_x$. Under the two no-impulse modes, survival rises steeply near $c_x=0.4$ and levels off
above about $0.5$. The designed minimum RN gap falls below the keep-out
for $c_x<0.35$. For the kick fault, survival continues to improve toward
isotropy: $0.71$ at $c_x=0.7$ versus $0.77$ at
$c_x=1$, favoring greater inclination spread. 

Inclination cost increases with the fleet mean of
$|\delta i|=\sqrt{c_x^2\delta e_x^2+\delta e_y^2}$.
Survival changes much more sharply than cost, leaving a
candidate band of roughly $c_x=0.4$ to $0.7$ for the
no-impulse modes. At $c_x=0.7$, paired trials give $1.5$ percentage
points higher held-attitude survival than the isotropic
grid (95\% confidence interval: $0.7$--$2.4$ points),
with uncontrolled-attitude survival within $0.1$ point.
The measured fleet-mean inclination cost is approximately
$13\%$ lower than the isotropic cost, but the estimated
worst-satellite budget is unchanged
(Table~\ref{tab:total_budget}).

The node ladder provides a separate alternative.
It places separation in the node component, with no
$\delta i_x$ or nodal shear at epoch. Its $\mathbf{\delta e}$
and $\mathbf{\delta i}$ patterns are parallel at $605$~m RMS, so
they still require co-rotation. This is the largest spread
among the designs and requires $31.7$~m/s/y, $1.7$ times the isotropic grid. Without imposed co-rotation, the ladder has zero
inclination-maintenance demand in the retained secular model. The ladder provides full pairwise
radial--normal protection and about five times the $c_x=0.7$ held-attitude $t_{0.90}$ (19.85 versus 3.7 d) but changes capacity
and topology. 

At $100$~m spacing over $\pm1$~km, it
holds $21$ satellites with a worst-case $20$ hops against
the disc's $14$. Matching the grid's count would require
$\pm4$~km and $80$ hops. Its smaller count also contributes
to survival: $210$ pairs against the grid's $3240$.
Whether the compute architecture and optical links can
support this chain remains unresolved. It may suit a
mission that accepts a linear topology, but has not been
demonstrated at equal capacity.

Selection among these options depends on the
requirements: satisfying keep-out and detection-plus-recovery latency at the
required survival probability. Until those are specified, the sweep
identifies a trade frontier rather than a single design.

\section{Future Work}
\label{sec:future}

Future work includes
extending control beyond the present capture behavior,
reducing the cost of preserving separation, and refining the system architecture.

\subsection{Control-law extension: orbit-lifted allocation}

The phase-by-phase law could be extended to allocate
effort over a full orbit. The orbit-integrated Gauss map separates
the effects of these acceleration components: constant along-track acceleration
controls $\delta a$, its first harmonic $\delta e$,
constant radial acceleration $\delta\lambda$, and
first-harmonic normal acceleration $\delta i$. A pitch-and-roll
profile with constant and first-harmonic terms could admit a stability proof; thrust supplies the normal pair.
Allocation then becomes a receding-horizon quadratic
program with an explicit propellant cost. This places
the current linear approximation to the braking curve at
the timescale where the $\delta a$--$\delta\lambda$
double-integrator structure holds (Sec.~\ref{sec:control}).

The present law's sliding surface is the
one-channel special case. 
A preliminary implementation gives a similar stationkeeping residual to the present law ($\approx2$~m) but takes longer to capture the tested initial displacement. Adding the second harmonic reduces the residual by $0.8$~m. Further work must connect local quadratic regulation with far-field switching and account for fleet coupling through the shared centroid. This formulation is not used in the results above.

\subsection{Geometry and reference extensions}

Geometry offers another way to change the cost. For static co-rotating
lattices, packing relates the fleet mean $|\delta i|$,
which sets cost, to the smallest pairwise gap, which
provides protection. For the isotropic construction, that gap equals the
minimum distance between lattice points. Hexagonal packing would reduce mean radius by
about $7\%$ in the bulk limit. A tight core with a loose
rim changes both mean radius and minimum gap. Allowing
the pattern angle to drift before correcting it impulsively
could save a few percent, but could risk safety.

We could also divide the work of maintaining
pattern alignment between thrust and attitude. Thrust
could rotate $\delta i$ through a fraction $p$ of the apsidal
angle, while attitude turns $\delta e$ back through the
remainder.

\subsection{System-level validation}

The orbital computing cluster concept has many remaining open questions.

\noindent \textbf{Optical links.} This analysis enforces no
link-range or footprint constraint. The line of sight between neighbors traces a $\pm30^\circ$ cone once per
orbit (not including attitude adjustment) about a mean direction that turns once per apsidal
period. Co-rotation keeps the cone constant, but the
terminals still need the corresponding gimbal coverage.
We have not modeled gimbal tracking rates, Sun exclusion, a yaw
schedule to absorb the slow term, or the simultaneous
terminals needed for the assumed degree.

\noindent \textbf{Network topology.} The node ladder is an example of a geometry that differs from the assumed 2D grid. Feasibility of such geometries depends on whether the AI compute architecture can operate over its linear network, a question beyond this orbital and control analysis.

\noindent \textbf{Fault detection in the shared reference.} A
corrupted state biases every other satellite's reference
by its error divided by fleet size; the present estimator
does not reject it. Optical-terminal bearings could help by comparing geometry, independent of the
navigation solution, and allow neighbors to detect a
disagreement between reported and observed position.

\noindent \textbf{Momentum control.} Propulsive unloading could reasonably be done with one burn per day, and a dual-use mounting of the normal thrusters, with a lever arm to absorb it as a byproduct of the inclination burns, is a possibility.

\noindent \textbf{Duration.} The longest closed-loop propagation
is $600$~orbits and the longest passive coast thirty days,
while the mission lasts years. The residual is bounded
over the tested horizon (Sec.~\ref{sec:stability}), but
a representative one-year propagation is needed to check
longer-term behavior.

Further work is needed on emergency avoidance, slot replacement after a vehicle loss, the structural response of a $16.75\times3$~m panel, thermal rejection, and the economic case. None is addressed here.

\section{Conclusion}

An $81$-satellite cluster at $100$~m spacing is maintained at
approximately $3$~m RMS slot error across nine simulated operating
conditions, using attitude-modulated forces in-plane and scheduled
normal thrust for inclination. The $600$-orbit case shows no secular
growth in the residual.

Co-rotation keeps the eccentricity and inclination patterns aligned,
preserving their designed radial--normal separation at about twice
the inclination-maintenance cost of holding inclination fixed.
For the isotropic grid, the combined three-year propellant
budget, including the adopted allowances, is $1.5$--$3.1$~kg
per satellite at the fleet mean and $2.3$--$3.8$~kg at the
rim across fixed solar-minimum to solar-maximum scenarios.

Geometry determines how long the formation can tolerate
control loss. At the $35$~m keep-out, the isotropic grid provides
$1.6$~days at $99\%$ survival after a single held-attitude failure,
compared with $6.5$~days under fleet-wide loss. The $c_x=0.7$ grid retains comparable five-day survival
under no-impulse failures with approximately $13\%$ less
inclination propellant than the isotropic grid. The node ladder extends survival but reduces
capacity and changes network topology.

These results make formation geometry and the tracked
reference part of the control design. A close-proximity computing
cluster must both hold its assigned slots and leave enough time to
respond when a vehicle stops doing so.

\appendix
\section{Force models}
\label{app:forces}

\noindent The controller uses a free-molecular flat plate of area $A$ at finite
speed ratio~\cite{storch}. The Monte Carlo model includes facets, while closed-loop tests also assumed a flat plate. For a corotating atmosphere, define the wind velocity relative
to the spacecraft by $\mathbf w=\boldsymbol\Omega_E\times\mathbf r-\mathbf v$,
where $\boldsymbol\Omega_E$ is Earth's rotation vector and
$\mathbf r,\mathbf v$ are inertial position and velocity.
Let $V=\|\mathbf w\|$, $\hat w=\mathbf w/V$,
$\hat n_w=\operatorname{sgn}(\hat n\cdot\hat w)\hat n$, and
$s=|\hat n\cdot\hat w|=\sin\gamma$, where $\gamma$ is the
nonnegative incidence angle from the plate plane. Then
\[
\mathbf F=\tfrac12\rho V^2 A(A_w\hat w+A_n\hat n_w).
\]
With $x=Ss$ and $f(x)=e^{-x^2}/\sqrt\pi+x\,\mathrm{erf}\,x$,
\begin{align*}
A_w&=\tfrac{C_t}{S}f(x),\\
A_n&=\tfrac{C_n}{2S^2}\mathrm{erf}\,x
 +\tfrac{C_n-C_t}{S}s f(x)+Ks.
\end{align*} 
\noindent The allocator's candidate generator replaces $f$ by a three-piece
$C^{1}$ quadratic $\hat f(x)=a+bx+gx^{2}$. The nine coefficients are determined
rather than fitted: $\hat f$ matches $f$ in value and slope at $x=0$, the third
piece is exactly $x$ so the hyperthermal limit is recovered exactly, and value
and slope are continuous at both knots. Only the knots are chosen, at $0.675$
and $1.575$, minimizing the maximum relative error, which is $0.555\%$ at
$x=0.822$.
\begin{center}\small
\begin{tabular}{@{}lrrr@{}}
\toprule
interval & $a$ & $b$ & $g$ \\
\midrule
$[0,\,0.675)$      & $0.56418958$ & $0$          & $0.52751127$ \\
$[0.675,\,1.575)$  & $0.39670677$ & $0.49624537$ & $0.15992210$ \\
$[1.575,\,\infty)$ & $0$          & $1$          & $0$ \\
\bottomrule
\end{tabular}
\end{center}

Accommodation $(\sigma_n,\sigma_t)=(0.90,0.80)$~\cite{traub-acc}
gives $C_n=2(2-\sigma_n)=2.2$ and $C_t=2\sigma_t=1.6$.
For $m_a=16$~u and $T_{\rm wall}=300$~K,
$V_w=\sqrt{\pi kT_{\rm wall}/(2m_a)}=494.85$~m/s and
$K=2\sigma_n V_w/V=0.11717$ at $V=7602$~m/s.
The speed ratio $S=V/\sqrt{2kT_\infty/m_a}$ follows solar
level; at $T_\infty=900$~K, $S=7.86$ and broadside $C_D=2.335$.

RTN satisfies $\hat R\times\hat T=\hat N$, with ram along
$+\hat T$. At $I=98^\circ$, corotation adds a headwind,
raising $V$ above orbital speed, and tilts the wind away from
$-\hat T$ by up to $\pm3.8^\circ$ about the radial axis.
Its changing sign contributes to pitch-response asymmetry.

\noindent Solar radiation pressure uses the standard specular--diffuse
flat-plate model~\cite{montenbruck-gill} (per facet in Monte Carlo),
{\small
\begin{equation*}
\mathbf a_{\rm srp}=-\frac{P_0A}{m}\cos\theta
\left[(1-r_s)\,\hat{\mathbf s}
+\Bigl(2r_s\cos\theta+\tfrac23 r_d\Bigr)\hat{\mathbf n}\right],
\end{equation*}
}
with $\cos\theta=\hat{\mathbf n}\cdot\hat{\mathbf s}$, $\hat{\mathbf s}$ from satellite to Sun, and $\hat{\mathbf n}$ the sunward
plate normal; the force is zero when $\cos\theta\le0$ or the satellite is in eclipse (cylindrical Earth shadow). Here
$P_0=4.54\times10^{-6}$~N/m$^2$, $r_s=0.05$, $r_d=0.07$.


\begin{thebibliography}{9}\small\setlength{\itemsep}{0pt}
\bibitem{concept} Ag\"uera y Arcas, B., Beals, T., Biggs,
M., Bloom, J.~V., Fischbacher, T., Gromov, K., K\"oster, U.,
Pravahan, R., and Manyika, J., ``Towards a Future
Space-Based, Highly Scalable AI Infrastructure System
Design,'' 2025, arXiv:2511.19468.

\bibitem{leonard} Leonard, C.~L., Hollister, W.~M., and
Bergmann, E.~V., ``Orbital Formationkeeping with
Differential Drag,'' \emph{Journal of Guidance, Control,
and Dynamics}, Vol.~12, No.~1, 1989, pp.~108--113,
doi:10.2514/3.20374.

\bibitem{damico} D'Amico, S., and Montenbruck, O.,
``Proximity Operations of Formation-Flying Spacecraft Using
an Eccentricity/Inclination Vector Separation,''
\emph{Journal of Guidance, Control, and Dynamics}, Vol.~29,
No.~3, 2006, pp.~554--563.

\bibitem{koenig} Koenig, A.~W., Guffanti, T., and D'Amico,
S., ``New State Transition Matrices for Spacecraft Relative
Motion in Perturbed Orbits,'' \emph{Journal of Guidance,
Control, and Dynamics}, Vol.~40, No.~7, 2017,
pp.~1749--1768.

\bibitem{traub} Traub, C., Fasoulas, S., and Herdrich, G.,
``A Planning Tool for Optimal Three-Dimensional Formation
Flight Maneuvers of Satellites in VLEO Using Aerodynamic
Lift and Drag via Yaw Angle Deviations,'' \emph{Acta
Astronautica}, Vol.~198, 2022, pp.~135--151.

\bibitem{traub-acc} Traub, C., Herdrich, G., and Fasoulas,
S., ``Influence of Energy Accommodation on a Robust
Spacecraft Rendezvous Maneuver Using Differential
Aerodynamic Forces,'' \emph{CEAS Space Journal}, Vol.~12,
2020, pp.~43--63, doi:10.1007/s12567-019-00258-8.

\bibitem{benyaacov} Ben-Yaacov, O., and Gurfil, P.,
``Long-Term Cluster Flight of Multiple Satellites Using
Differential Drag,'' \emph{Journal of Guidance, Control, and
Dynamics}, Vol.~36, No.~6, 2013,
doi:10.2514/1.61496.

\bibitem{bevilacqua} Bevilacqua, R., and Romano, M.,
``Rendezvous Maneuvers of Multiple Spacecraft Using
Differential Drag Under $J_2$ Perturbation,'' \emph{Journal
of Guidance, Control, and Dynamics}, Vol.~31, No.~6, 2008,
doi:10.2514/1.36362.

\bibitem{spacex-thruster} SpaceX,
``Argon Hall Thruster Tech Specs,''
Twitter (now X), 26 February 2023.
\url{https://x.com/SpaceX/status/1629948869239873538}.

\bibitem{koenig18} Koenig, A.~W., and D'Amico, S., ``Robust
and Safe N-Spacecraft Swarming in Perturbed Near-Circular
Orbits,'' \emph{Journal of Guidance, Control, and Dynamics},
Vol.~41, No.~8, 2018, pp.~1643--1662.

\bibitem{bocci} Bocci, A., Corona-S\'anchez, J.~J., and
Kristiansen, R., ``Leader--Follower Formation Control Using
Differential Drag and Effective Surface Regulation,''
arXiv:2604.18140, 2026; presented as AAS 25-661, AAS/AIAA
Astrodynamics Specialist Conference, Boston, 2025.

\bibitem{smad} Wertz, J.~R., Everett, D.~F., and Puschell,
J.~J. (eds.), \emph{Space Mission Engineering: The New
SMAD}, Microcosm Press, Hawthorne, CA, 2011, Table~I-1,
``Earth Satellite Parameters,'' p.~1031.

\bibitem{montenbruck-gill}
O.~Montenbruck and E.~Gill, \emph{Satellite Orbits: Models, Methods and
Applications}, Springer, Berlin, 2000, Sec.~3.4.

\bibitem{storch} Storch, J.~A., ``Aerodynamic Disturbances
on Spacecraft in Free-Molecular Flow,'' Aerospace Corp.\
Rept.\ TR-2003(3397)-1, 2002.

\bibitem{olfati} Olfati-Saber, R., Fax, J.~A., and Murray,
R.~M., ``Consensus and Cooperation in Networked Multi-Agent
Systems,'' \emph{Proceedings of the IEEE}, Vol.~95, No.~1,
2007, pp.~215--233.

\bibitem{planet} Vittaldev, V., Leung, S., Beukelaers, V.,
Stepan, L., and Zimmerman, R., ``Constellation Phasing with
Differential Drag on Planet Labs Satellites,'' \emph{Journal
of Spacecraft and Rockets}, Vol.~55, No.~2, 2018,
pp.~473--483, doi:10.2514/1.A33927.

\bibitem{lara2015} Lara, M., ``Efficient Formulation of the Periodic
Corrections in Brouwer's Gravity Solution,'' \emph{Mathematical Problems
in Engineering}, Vol.~2015, Article~980652, 2015, pp.~1--9.
doi:10.1155/2015/980652.

\end{thebibliography}
\end{document}